\documentclass[prl,twocolumn,superscriptaddress,showpacs]{revtex4}
\usepackage{graphicx}% Include figure files
\usepackage{epsfig}% Include figure files

\usepackage{amssymb,amsmath,amsfonts,hyperref}
\usepackage{wasysym}
\usepackage{latexsym}

\newcommand{\nn}{\nonumber}

\newcommand{\be}{\begin{eqnarray}}
\newcommand{\ee}{\end{eqnarray}}

\begin{document}
\preprint{APS/123-QED}
%\draft
%\twocolumn[\hsize\textwidth\columnwidth\hsize\csname @twocolumnfalse\endcsname
\title{Collinear ordering of easy-axis triangular lattice antiferromagnets}
\author{Arnab Sen}
\affiliation{{\small Department of Theoretical Physics,
Tata Institute of Fundamental Research,
Homi Bhabha Road, Mumbai 400005, India}}
\author{Fa Wang}
\affiliation{{\small Department of Physics, University of California, Berkeley, CA 74720}}
\author{Kedar Damle}
\affiliation{{\small Department of Theoretical Physics,
Tata Institute of Fundamental Research,
Homi Bhabha Road, Mumbai 400005, India}}
\affiliation{{\small Physics Department, Indian Institute of Technology Bombay, Mumbai 400076, India}}

\vspace{-2 cm}
\date{\today}
\begin{abstract}
Antiferromagnetically coupled moments on the frustrated triangular lattice typically order in
a coplanar state at low temperature. Here, we demonstrate that the presence of not-very-large easy axis single ion anisotropy
leads to an interesting orientationally ordered collinear state in triangular lattice antiferromagnets with moments $S \ge 3/2$.  This ordered state
breaks the symmetry of $\pi/3$ rotations about a lattice site, while
leaving intact the translational symmetry of the lattice. 
\end{abstract}

\pacs{75.10.Jm, 75.25.+z}
\vskip2pc

\maketitle
%\section{Introduction}

{\em Introduction:}
Magnetic ions in insulating solids often interact through `exchange'
couplings
between nearest neighbours. In some cases, these exchange couplings
compete with each other due to the geometry of the lattice. Such
`geometrically frustrated' magnets typically
have a broad cooperative paramagnetic regime~\cite{Moessnerreview} for
a range of temperatures below the exchange energy scale. The behaviour
of the system in this regime is controlled by the
manner in which the spins explore the multitude
of classical ground states that arise from the geometric
frustration. This gives way at very low temperature to a variety of
novel phases~\cite{Misguich} arising from quantum effects and
sub-leading
interactions.

Antiferromagnets on the triangular lattice provide particularly well-studied examples of magnetic frustration. When spins on the triangular lattice are coupled by nearest neighbour `Heisenberg exchange' coupling that respects the symmetry of spin-rotations, they order in a coplanar 120$^\circ$ ordered state for large spin $S$, and the weight of accumulated evidence~\cite{Capriotti99} suggests that this ordering of Heisenberg antiferromagnets on the triangular lattice survives the effects of quantum fluctuations and persists down to $S=1/2$. In the opposite {\em Ising} limit,
when the exchange couplings only couple one component of the spins, there are of course no quantum fluctuations; however, as was shown long ago by Wannier~\cite{Wannier} in the $S=1/2$ case, the system remains in a collinear paramagnetic state all the way down to $T=0$ due to the geometric frustration of the triangular lattice.

Here, our focus is on the role of quantum effects in driving the large spin ($S \ge 3/2$) triangular lattice Heisenberg antiferromagnets with relatively strong easy-axis single ion anisotropy to an ordered collinear state. As we demonstrate below, the leading effects of virtual quantum transitions induced by the transverse components of the exchange coupling pick out an unusual `orientationally ordered' collinear state for all $S \ge 3/2$ whenever the easy axis single ion anisotropy $D$ dominates over the isotropic nearest neighbour exchange $J$.

\begin{figure}
\includegraphics[width=\hsize]{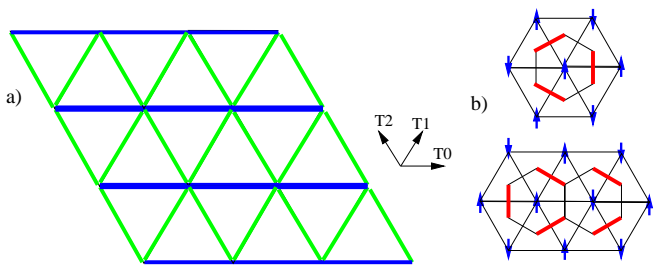}
      \caption{(color online). a) Orientationally ordered state on the
	triangular lattice. The average Ising exchange energy of the
	blue (dark) bonds is lower than that of the green (light)
	bonds. b) The representation of
minimally frustrated Ising configurations in terms of dimers (dark bonds) on the dual honeycomb net.}
      \label{brokensymmetry}
  \end{figure}

In this ordered state which breaks the symmetry of $\pi/3$ rotations
about a lattice site, but does not  break lattice translation
symmetry (Fig~\ref{brokensymmetry} a), the total magnetization along
the $z$ axis remains zero for small applied field $B$ along the easy axis, and the system is thus expected to have a $m=0$ plateau. This is in sharp contrast to $S=1$ magnets with similar easy axis anisotropy---as was pointed out in Ref~\cite{Damle_Senthil}, the original results of Refs.~\cite{Heidarian_Damle,Melko_etal,Wessel_Troyer} imply that $S=1$ antiferromagnets with strong easy axis anisotropy on the triangular lattice order in a state that has coexisiting three-sublattice spin-density wave order for the $z$ component of the spin, as well as spin-nematic order in the transverse components of the spins.

As we detail below, these results are expected to be of potential
relevance to some members of a class of interesting magnets~\cite{Flahaut_etal,Hardy_etal1, Hardy_etal2,Vajenine_etal_etc,Mohapatra_etal} in which one set of magnetic ions have an unusual trigonal prismatic environment that typically leads to significant single-ion anisotropy effects in the low energy spin Hamiltonian~\cite{Hempel_Miller,Kageyama_etal,Wu_etal,Burnus_etal}.

{\em Model and Motivation:}  We focus on the effects of a large easy axis single-ion anisotropy on triangular lattice antiferromagnets with Hamiltonian
\be
H_D = J\sum_{\langle ij \rangle}\vec{S_i}\cdot\vec{S_j} - D\sum_{i}(S^z_{i})^2 - B\sum_{i}S^z_{i},
\ee
where $J > 0$ denotes the nearest neighbour antiferromagnetic spin exchange interaction between the $S \ge 3/2$ ions, $D$ is the strength of the easy axis single-ion anisotropy along the $z$ axis, and $B$ represents the external field along the easy axis.

Because of the layered nature of triangular magnets, uniaxial
single-ion anisotropy that picks out a common easy axis is a generic
possibility in such systems. For instance, this possibility is
frequently realized on the A
site~\cite{Hempel_Miller,Kageyama_etal,Wu_etal,Burnus_etal}. in the
A$^{'}_3$ABO$_3$ family of compounds (with A$^{'}$ $=$ Ca, Sr, or
Ba)~\cite{Flahaut_etal,Hardy_etal1,Hardy_etal2,Vajenine_etal_etc,Mohapatra_etal}
which consist of a triangular array of chains of alternating `A' and
`B' type sites that can host a wide variety of magnetic or
non-magnetic cations~\cite{Hardy_etal1} with typically
antiferromagnetic inter-chain coupling and intra-chain coupling of
variable sign and magnitude. Although our two dimensional model does
not apply directly in cases with large intra-chain coupling (such as
Ca$_3$Co$_2$O$_6$), it is
nevertheless expected to be a good starting point for other compounds
of this type in which the B site is non-magnetic, and the coupling
between successive high-spin A sites along the chain is concomitantly weak compared to the in-plane exchange coupling.

Returning to our model Hamiltonian, we note that the anisotropy term can induce a collinear state for not-very-large values of $D/J$ due to the frustrated nature of the exchange (for instance, in the classical large $S$ limit, the system prefers a collinear state for $D > 3J/2$). The detailed nature of the collinear ordering is then controlled by the leading effects of quantum fluctuations induced by the exchange coupling $J$. In order to understand these effects, we develop a systematic large $D/J$ expansion for the leading order effective Hamiltonian that governs the effects of these fluctuations---this perturbative analysis is expected to yield reliable results whenever the easy axis anisotropy forces the spins to be collinear. An analysis of the low temperature properties of this effective Hamiltonian then yields detailed predictions for the collinear ordering selected in this regime.

{\em Method:} We begin by splitting the Hamiltonian as $H = H_0 + H_1$ with $H_0 =  - D\sum_{i}(S^z_{i})^2$ and $H_1 = J\sum_{\langle ij \rangle}\vec{S}_{i} \cdot \vec{S}_{j}$. Clearly, the ground state manifold of $H_0$ at large $D$ is obtained by requiring all spins to be maximally polarized along the $z$ axis and have $S^z_i = \sigma_i S$, with $\sigma_i = \pm 1$. The low energy physics in this {\em Ising subspace} is then controlled by the leading perturbative effects of the exchange coupling $J$. These perturbative effects can be represented as an effective Hamiltonian ${\cal H}$ acting within this low energy subspace. Using standard degenerate perturbation theory and including terms up to ${\mathcal{O}}(J^3/D^2)$, we find (upto unimportant additive constants) for $S=3/2$:
\be
\mathcal{H} &=& J_1 \sum_{\langle ij \rangle}\sigma_i \sigma_j -
J_2 \sum_{\langle ij \rangle} \frac{1-\sigma_i
  \sigma_j}{2}(\sigma_i H_i + \sigma_j H_j) \nn \\
 &+& J_3
\sum_{\langle ij \rangle} (\sigma_i^{+}\sigma^{-} + \mathrm{h.c.}) 
\ee
where $J_1 = \frac{9J}{4}$, $J_2 = \frac{27J^3}{64D^2}$, and $J_3 = \frac{9J^3}{64D^2}$ for $S = 3/2$, while the {\em exchange field} $H_i \equiv \Gamma_{ij}\sigma_j$ with $\Gamma_{ij}=1$ for nearest neighbours and zero otherwise.

In the above, the first term corresponds to the leading effect of the
$z$ component of the spin exchange, while the second term arises from
the effects of {\em virtual} quantum transitions of pairs of
anti-aligned spins out of the low energy Ising subspace: such virtual
transitions cost differing amounts of energy depending on the local
environment of the fluctuating pair, and these differences lead to
different `zero-point' energies which are encoded in this `potential
energy' term. Both these {\em diagonal} terms generalize
straightforwardly to arbitrary $S>3/2$ and we find $J_1(S) = JS^2$ and
$J_2(S) = \frac{S^3J^3}{4D^2(2S-1)^2}$. In contrast the presence of
the off-diagonal term (representing real quantum transitions) at order
${\mathcal O}(J^3/D^2)$ is special to $S=3/2$, and it is easy to see
that the effects of real quantum transitions are in general of order
${\mathcal O} (J^{2S}/D^{2S-1})$ and can be safely ignored for
$S>3/2$. Moreover, even for $S=3/2$, the $J_3$ term has a
significantly smaller numerical coefficient compared to the other
subleading term $J_2$, and we therefore expect the effects of the
potential energy term $J_2$ to dominate over the off-diagonal $J_3$
term for $S=3/2$ magnets as well.\begin{figure}
\includegraphics[width=\hsize]{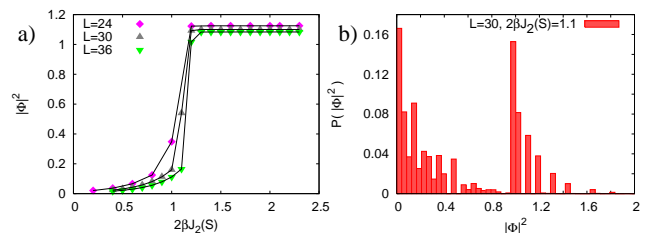}
      \caption{(color online). a) Orientational order parameter $\Phi$ as a function of the inverse temperature. b) The double peak in the histogram of $\Phi$ provides a clear signature of a first order jump in $\Phi$ at the transition.}
      \label{transition}
  \end{figure}

Thus, the low temperature behaviour is well-described by an effective
{\em classical triangular lattice Ising model} written in terms of
Ising pseudospin variables $\sigma$ that interact with a leading
nearest-neighbour antiferromagnetic coupling $J_1$, and a sub-leading
multi-spin interaction $J_2$ arising from virtual quantum
transitions. At low temperature, the leading term constrains the
system to lie within the macroscopically degenerate manifold of {\em
  minimally frustrated} states (in which every triangle has precisely
one {\em frustrated} bond corresponding to a pair of ferromagnetically
aligned
$\sigma$) of the classical triangular lattice Ising antiferromagnet. Our focus here is therefore on the possible effects of the subleading multispin interaction term $J_2$ acting within this degenerate subspace.

Further analysis of the effects of this unfamiliar multi-spin interaction on the low temperature physics is facilitated by noting that the minimally frustrated states of the classical Ising antiferromagnet may be represented as dimer covers of the dual honeycomb lattice, with a dimer placed on every dual lattice link that crosses a {\em frustrated bond} of the physical lattice (Fig~\ref{brokensymmetry} b). The multispin interaction $J_2$ acting within this space of states can then be written in this language as an interaction term $\mathcal{V}$ between the dimers.

To see this, we first note  that this term only acts on {\em
  unfrustrated bonds} of a pseudospin configuration. Now, since the number of unfrustrated bonds in any minimally frustrated configuration
is a constant equal to $n_t$, the number of triangles in the system, we may rewrite the potential energy $\mathcal{V}$ as:
$\mathcal{V} = -2J_2(S)n_t \sum_{n=2}^{5}nf_n$
where $f_n$ denotes the fraction of unfrustrated bonds with $n$  dimers on the perimeter of the corresponding double-hexagon of the dual honeycomb lattice (Fig~\ref{brokensymmetry} b).

{\em Semiclassics: } In order to test this general result for the potential energy term $\mathcal{V}$, we have also performed a semiclassical expansion in the large $S$ limit and calculated the effective Hamiltonian (restricted, as above, to the minimally frustrated subspace of the classical Ising antiferromagnet) at leading ${\cal O}(S)$ order using the procedure outlined in Ref~\cite{Hizi_Henley}. Expanding this large-$S$ semiclassical result to leading order in the physical
small parameter $J/D$, we obtain (apart from unimportant additive constants)
\be
{\cal H}_{SC} &=& \frac{J^3S}{32D^2} \sum_i  H_i^2
\ee
This result is, at first sight, quite different from the outcome of
our perturbative analysis. However, it is not difficult to see that
${\cal H}_{SC}$ is in fact equivalent to the large $S$ limit of ${\cal H}$. To see this, we first note that ${\cal V}/2J_2(S)$ just counts the total number of dimers on the perimeters of double hexagons of the dual lattice. On the other hand, the semiclassical potential energy ${\cal H}_{SC}$ derived above can be written as ${\cal H}_{SC} = 2J_2(S) n_h \sum_{m=0}^{3} m^2g_m$  where $n_h \equiv n_t/2$ is the total number of hexagons in the dual lattice, and $g_m$ denotes the fraction of hexagons with $m$ dimers on its perimeter.

Now, each single hexagon with $m$ dimers on its perimeter can be part of $6-m$ such double hexagons corresponding to unfrustrated bonds, and these $m$ dimers will thus contribute $6-m$ times in this sum. As a result, ${\cal V}$ can equally well be written as ${\cal V} =  -2J_2(S)n_h\sum_{m=0}^{3} m(6-m) g_m$, Simplifying and taking the large-$S$ limit, we immediately obtain (apart from an additive constant) the alternate single-hexagon form of the potential energy  that emerged from the semiclassical analysis.

{\em Low temperature state: }
Since the average number of dimers on the perimeter of a hexagon is $2$, it is clear from the single hexagon form of ${\cal V}$ that the minimum potential energy is obtained for all configurations with $g_2=1$, {\em i.e.} for configurations with all hexagons having precisely two dimers on their perimeter. To proceed further and understand the nature of the low temperature state chosen, it is necessary to efficiently simulate the corresponding interacting dimer model at low temperature.

In order to do this, we employ a generalization~\cite{Sen_Damle_Moessner} of the procedure of
Refs~\cite{Sandvik_Moessner,Alet_etal}  and work with the single hexagon form of ${\cal V}$. Our conclusions from this numerical study are readily stated: We find that below a critical temperature $T_c \approx 1.67 J_2(S)$, the system orders into a {\em orientationally ordered state} in which the mean Ising exchange energy on a link of the triangular lattice depends on its orientation, but not its position. [We note parenthetically that a very similar ordered state is also realized in another interacting dimer model studied recently in an entirely different context~\cite{Sen_Damle_Vishwanath}.]

This is seen in the behaviour of the orientational order parameter  $\Phi=
\sum_p -B_p e^{2p\pi i/3}$, where $B_p$ denotes the average of the Ising exchange energy $\sigma_i \sigma_{j}$ on all links $\langle i j \rangle$ of the $p^{th}$ orientation ($p=0,1,2$--Fig~\ref{brokensymmetry} a) on the triangular lattice (Fig~\ref{transition} a). From the double peak nature of the histogram of the order parameter at the transition, we see that the transition has a first order character (Fig~\ref{transition} b).

 \begin{figure}
\includegraphics[width=\hsize]{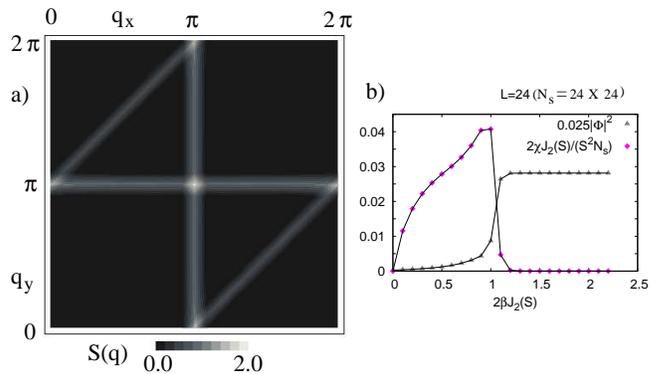}
      \caption{(color online). a) Contour plot of the
	pseudo-spin structure factor $S(\vec{q})$ for size $L = 24$ at $2\beta
	J_2(S) = 2.2$ ($q_x$ and $q_y$ are projections of the
	wavevector $\vec{q}$ along lattice directions $T_0$ and $T_1$ (Fig~\ref{brokensymmetry})). The Bragg lines with enhanced signal are the
	signature of the orientational order. b) The magnetic
	susceptibility for fields along the easy axis. Note the sharp drop in susceptibility when the system enters the orientationally ordered state. }
      \label{bragglines}
  \end{figure}

In such an orientationally ordered state, the Ising pseudospins are
antiferromagnetically arranged in parallel rows oriented along one of
the three principal directions of the triangular lattice. As each
such row can be in one of two internal states corresponding to
the two antiferromagnetic arrangements of $\sigma$ on that row, the
question then arises: Are these internal states (that may be
represented by a `parity' variable that takes two values) 
locked into some specific ordered pattern, or are they
fluctuating
randomly? To answer this question, we monitor the structure
factor of the pseudospins $\sigma$ (which corresponds to
experimentally measurable leading $zz$ component of the structure
factor of physical spins $\vec{S}$)
and note that if these internal parity variables had long-range
order at some wavevector, one would expect to see a corresponding
Bragg peak
 in the pseudo-spin structure factor, while the absence of any
 observed Bragg peaks would imply either a
 random {\em glassy} pattern of parities or rapidly fluctuating
 parities.

As is clear from Fig~\ref{bragglines} a, there is no long range order in the
internal states of the antiferromagnetic rows. Indeed, in our
numerics, we find that the antiferromagnetic rows fluctuate freely
between their two allowed internal states. However, these fluctuations
are only possible because
the algorithm uses very efficient non-local multi-spin moves to
rapidly equilibriate the system with the correct equilibrium Gibbs
distribution. With realistic single-spin flip or spin-exchange
dynamics, we find that the antiferromagnetic rows freeze into a random glassy
pattern of internal parity states as the equilibrium transition
temperature is approached. Such glassy behaviour due to the presence
of a diverging time scale associated with the formation of extended
structures has been discussed at length in earlier theoretical studies~\cite{Das_Kondev_Chakraborty},
and our results provide another example of such glassy behaviour in
a disorder-free system. [We also note parenthetically that a
  similar ordered state, but with the row parities `ferromagnetically'
  arranged, is realized in the spin-$1$ triangular lattice  antiferromagnet $\mathrm{AgNiO}_2$ but the underlying mechanism is likely to be quite different~\cite{wawrzynska}.]

Another interesting aspect of this orientationally ordered state
arises from the fact that the total magnetization $\langle M_{\mathrm tot} \rangle$ as well
as the uniform magnetic susceptibility $\chi$ to a magnetic field
along the $z$ axis are exponentially small deep inside the ordered state (see Fig~\ref{bragglines} b).
As this ordered state is stable to small magnetic fields for which
the potential energy ${\cal V}$ dominates over the Zeeman energy, this implies
the presence of low temperature {\em zero magnetization plateau} that extends for a range of
magnetic
fields $0 < |B| < B_c \sim J^3/D^2$. [For $|B| > B_c$, the Zeeman energy gain
dominates over the fluctuation induced potential energy, and the
system is expected to order in the well-known three-sublattice
ordered one-third magnetization state of the triangular lattice
Ising antiferromagnet.]

{\em Discussion:}
 As mentioned earlier, our predictions for an orientationally ordered
 state
at low temperature are expected to be of relevance
 to other members of the Ca$_3$Co$_2$O$_3$ family
 of compounds in which the octahedral site is non-magnetic, resulting in weak (in comparison to the in-plane antiferromagnetic exchange coupling) coupling
between successive high-spin magnetic ions on the trigonal prismatic
 sites along the chain. The orientational ordering preserves the translational
 symmetry of the triangular lattice, but breaks the rotational
 symmetry $\pi/3$ rotations about a lattice site, and is expected to
give rise to
a characteristic set of high intensity lines in the $zz$ component
of the spin structure factor, which can be probed in neutron
 scattering
experiments. Furthermore, the ordered state
is also associated with a zero-magnetization plateau in the magnetization
response to a field applied along the easy axis, while the ordering
transition is expected to give rise to a sudden drop in the magnetic susceptibility
$\chi$ as well as a prominent signature in the specific heat. In
 addition, the ordered state is also expected to display signatures of
slow glassy dynamics of the spins.
Given the large
variety of magnets that crystallize in this structure and the recent interest in this family of materials,
we hope that our work provides some impetus to experimentally identify
 examples of this interesting physics.

{\em Acknowledgements:} We acknowledge valuable discussions
with R.~Coldea and E.~Sampathkumaran, and especially thank
A.~Vishwanath for many valuable comments as well
as collaboration on closely related work. Computational resources of TIFR, and funding from LBNL DOE-504108 (FW),
and DST SR/S2/RJN-25/2006 (KD) are also gratefully acknowledged.

\end{document}